\documentclass{article}
\usepackage{spconf}

\usepackage{graphicx,xcolor}
\graphicspath{{../pdf/}{../jpeg/}}
\DeclareGraphicsExtensions{.pdf,.jpeg,.png}
\usepackage{amsmath}
\usepackage{amsfonts}
\usepackage{array}
\usepackage{amssymb}
\usepackage{empheq}
\usepackage{epstopdf}
\usepackage[linesnumbered,ruled]{algorithm2e}
\newcounter{MYtempeqncnt}
\usepackage{array}
\usepackage{caption}
\usepackage{subcaption}

\ninept
\begin{document}
\title{Distributed Network Caching via Dynamic Programming}
\name{Alireza Sadeghi$^\dagger$, Antonio G. Marques$^*$, and Georgios B. Giannakis$^\dagger$
	\thanks{{ The work in this paper has been supported by USA NSF grants 1423316, 1508993, 1514056, 1711471, and by the Spanish MINECO grants OMICROM (TEC2013-41604-R) and  KLINILYCS (TEC2016-75361-R).}}}
\address{$^\dagger$Digital Technology Center and Dept. of ECE, University of Minnesota, Minneapolis, USA \\
	$^*$Dept. of Signal Theory and Comms., King Juan Carlos University, Madrid, Spain}
\maketitle
\maketitle
\begin{abstract}
Next-generation communication networks are envisioned to extensively utilize storage-enabled caching units to alleviate unfavorable surges of data traffic by pro-actively storing anticipated highly popular contents across geographically distributed storage devices during off-peak periods. This resource pre-allocation is envisioned not only to improve network efficiency, but also to increase user satisfaction. In this context, the present paper designs optimal caching schemes for \textit{distributed caching} scenarios. In particular, we look at networks where a central node (base station) communicates with a number of ``regular'' nodes (users or pico base stations) equipped with \textit{local storage} infrastructure. Given the spatio-temporal dynamics of content popularities, and the decentralized nature of our setup, the problem boils down to select what, when and \textit{where} to cache. To address this problem, we define fetching and caching prices that vary across contents, time and space, and formulate a global optimization problem which aggregates the costs across those three domains. The resultant optimization is solved using decomposition and dynamic programming techniques, and a reduced-complexity algorithm is finally proposed. Preliminary simulations illustrating the behavior of our algorithm are finally presented. 
\end{abstract}
\section{Introduction}
%
The tremendous growth of data traffic  over wired and wireless networks  motivates the need for emerging technologies to meet the ever-increasing data demands~\cite{Role_Cache}. Recognized as an appealing solution is the so-called \textit{caching}, which refers to storing reusable popular contents across geographically distributed storage-enabled network entities~\cite{Role_Cache, Bastug}. The rationale here is to smoothen out the unfavorable shocks of peak traffic periods by pro-actively storing anticipated highly popular contents at these storage devices and during off-peak instances. This idea of resource pre-allocation is envisioned to achieve significant network resource savings in terms of e.g., energy, bandwidth, and cost. Which already comes with user satisfaction~\cite{Role_Cache, Bastug, Negin}.

Utilizing (stochastic) optimization approaches to enable content-agnostic storage-enabled network entities to pro-actively cache popular contents has quickly gained attention. Under {\it  static popularity} settings, a multi-armed bandit formulation is considered in~\cite{Gunduz}. The coded, and convexified version is studied in~\cite{Sengupta}. Context and trend aware learning approaches are provided in~\cite{Context, Trend} and a transfer learning one in~\cite{Transfer}. Belief propagation based distributed cooperative caching is investigated in~\cite{Beliefe}, and an ADMM based distributed approach in~\cite{ADMM}. Recently, by considering {\it dynamic popularities},  reinforcement learning approaches are utilized to cache contents~\cite{Sadeghi, Gunduz2, Sadeghi2}.

Existing caching approaches {accounting for dynamic popularities} focus on enabling a \textit{single cache} entity to decide (learn) what and when to cache. However, since caching is being carried out in a network level, to fully unleash its potentials a distributed approach is essential to efficiently utilize the globally available storage capacity over the entire network. Building on this observation, the present paper proposes optimal dynamic schemes to operate over a network where nodes, connected to a central controller, receive user requests and are equipped with local cache memories where they can store specific contents. In particular, the optimal cache-fetch decisions across caching entities will be found as the solution of a constrained optimization aimed at reducing an expected network cost aggregated across caching entities, contents and time instants. Since the caching decision at a given node not only affects the instantaneous cost but also the availability of the content in the next slots, the problem is indeed a dynamic programming  (DP) one.  DP algorithms typically involve functional estimation and, as a result, they are computationally expensive. A key aspect of this paper is to combine a marginalization and a decomposition approach to effectively reduce the dimensionality of the problem and, hence, limit the computational complexity. 
\section{Problem Formulation}
Consider a distributed caching system composed of a central node (CN) connected to a number of regular nodes, all devised with local storage devices. The CN can be a base station in a cellular network, or a gateway router in a content delivery network. The nodes are then user devices or access points, correspondingly. We consider $m = 1 \cdots M$ nodes,  $f = 1,2 \cdots F$ files, and $t = 1, 2, \cdots$ time instants. For convenience the CN, which is also connected to a cloud through a (congested) back-haul, is indexed by $m=0$. With this notation, at a given slot, every node receives user file requests as indicated by the binary variable $r_{t,m}^f$, where $r_{t,m}^f = 1$ denotes that during slot $t$ file $f$ was requested at node $m$, and $r_{t,m}^f = 0$ otherwise. Upon receiving a file request, the node forwards the request to the CN unless the file is already available in the cache of the node. If file $f$ is available at the CN, the content is served to node $m$. If not, the CN fetches required file either from the cloud or from one of the local nodes. Moreover, once the content has been delivered, all the nodes having the file must decide whether to keep it or to discard it. Decisions are made based on the state of the system as well as on the prices (costs) associated with caching and fetching, with the overall goal being to minimize the long-term aggregated cost. 
%
\subsection{Optimization variables and constraints}
To formalize the setting, let us define the binary \textit{fetching} variables ${\underline w}^f_{t,m}$, and $\overline w^f_{t,m}$,  as well as the binary \textit{storing} decision variable $a^f_{t,m}$. For the nodes, having ${\underline w}^f_{t,m} = 1$, implies fetching file $f$ at slot $t$ from node $m$ to the CN, while ${\underline w}^f_{t,m} = 0$ means ``not fetching''. Similarly $\overline w^f_{t,m}$ indicates if the CN fetches the file $f$ from node $m$ at time $t$.  To have a compact notation, fetching from the cloud to the CN is denoted as $w^f_{t,0} = \underline w^f_{t,0} = \overline w^f_{t,0}$. For all the nodes, the indicator $a^f_{t,m} = 1$ means that file $f$ will be stored at node $m$ at the end of slot $t$, and the binary storage state variable $s^f_{t,m}$ accounts for availability of file $f$ at node $m$ at the beginning of slot $t$. The availability of file $f$ directly depends on the preceding caching decision, thus the first set of constraints is
\begin{equation}
{\textrm {(C$1$)} }\quad s^f_{t,m} = a^f_{t-1, m},
\end{equation}   
which must hold $\forall f,t, m$. Having $r^f_{t,m} = 1$ and $s^f_{t,m} = 0$  necessitates requesting the file from the CN, which in turn requires fetching it either  from  the cloud or from the nodes having the content. This gives rise to the second set of constraints
\begin{equation}
\label{c2}
{\textrm {(C$2$)} }\quad r^f_{t,0} \le w^f_{t,0} + s^f_{t,0} + \sum \limits_{m = 1}^{M}  \left({\underline w}^f_{t,m} \; . \;  s^f_{t,m} \right).
\end{equation}
The product ${\underline w}^f_{t,m} .  s^f_{t,m}$ guarantees that fetching from a node is feasible only if the node has the file. 
Storing a file at a node necessitates to either the file being available locally or fetching it from the CN, giving rise to the following constraint 
\begin{equation}
{\textrm {(C$3$)} }\quad a^f_{t,m} \le s^f_{t,m} + \overline w^f_{t,m}, \; \forall m = 1 \cdots M.
\label{c3}
\end{equation}
Likewise, storing at the CN requires
\begin{equation}
{\textrm {(C$4$)} }\quad a^f_{t,0} \le s^f_{t,0} + w^f_{t,0} + \sum \limits_{m = 1}^{M} \underline w^f_{t,m} \; . \; s^f_{t,m} 
\label{c4}
\end{equation}
Finally, a node can fetch only if  
\begin{equation}
{\textrm {(C$5$)} }\quad \overline w ^f_{t,m} \le s^f_{t,0} + w^f_{t,0} + \sum \limits_{m' \ne m} \underline w^f_{t,m'} \; . \; s^f_{t, m'}.
\label{c5}
\end{equation}
\subsection{Prices and aggregated costs}
As assumed earlier, both caching and fetching are costly. For node $m$, at time $t$ and for file $f$, let $\rho^f_{t,m}$, ${\underline  \lambda}^f_{t,m}$, and $ {\overline \lambda}^f_{t,m}$ stand for costs of storing, fetching to, and fetching from the CN, respectively. Then, the overall incurred cost is  $c^f_t = \sum_{m=0}^{M} c^f_{t,m}$, where for regular nodes we have that 
\begin{equation}
c^f_{t,m} := \overline \lambda^f_{t,m} \overline w^f_{t,m} + \rho^f_{t,m}  a^f_{t,m}, \;\forall m = 1 \cdots M,
\label{cost}
\end{equation}
while for $m=0$ (the CN) we have 
\begin{equation}
	c^f_{t,0} := \lambda_{t,0} w^f_{t,0} + \rho^f_{t,0}  a^f_{t,0} +\sum \limits_{m = 1}^{M} \underline \lambda^f_{t,m} \underline w^f_{t,m},
\label{c_0}
\end{equation}
with $\lambda_{t, 0}$ being the cost of fetching from the cloud and $\rho_{t, 0}$ being the caching cost for the CN.
\subsection{Problem Formulation}
Let us collect all caching and fetching costs during slot $t$ for file $f$ at vectors $\pmb \rho^f_{t}$ and ${\pmb \lambda}^f_t = \left[ \underline {\pmb \lambda}^f_t, \overline {\pmb \lambda}^f_t\right]$, and define similarly the action vectors ${\pmb a}^f_{t}$ and ${\pmb w}^f_{t}=[\underline {\pmb w}_t, \overline {\pmb w}_t]^{\top}$ as well as the request vector $\pmb r_t^f$. The first step is to define the aggregated cost to be minimized. Since there is an inherent uncertainty in the state of the system, we consider the \textit{discounted} long-term incurred average cost
\begin{figure*}[th!]
	\vspace{.04 cm}
	\setcounter{MYtempeqncnt}{\value{equation}}
	\begin{align} 
	\label{long1} \left({\pmb w}_t^{\ast}, {\pmb a}_t^{\ast}\right)  :&= \!\!\!\!\!\underset{(\pmb w,\pmb a) {\textrm { s.t. (C1)-(C5)}} }{\arg\mathop{\min}} \left \{ {\mathbb E} \left[\underset{({\pmb w}_k,{\pmb a}_k) {\textrm { s.t. (C1)-(C5)}} }{\min} \left\{\sum \limits_{k=t}^{\infty} \gamma^{k-t} \left[c_k \left(\pmb a_k,\pmb w_k;\pmb \rho_k,\pmb \lambda_k\right) \Big | {\pmb a}_t = {\pmb a}, {\pmb w}_t = {\pmb w}, {\pmb s}_t = {\pmb s}_0, {\boldsymbol \theta_t = {\boldsymbol \theta}_0} \right]\right\}\right] \right\}\\ 
	\label{long2} & = \!\!\!\!\! \underset{(\pmb w,\pmb a) {\textrm { s.t. (C1)-(C5)}} }{\arg\mathop{\min}} \left\{c_t \left(\pmb a,\pmb w;\pmb \rho_t,\pmb \lambda_t\right)  +   \underset{\boldsymbol \theta_k}{\mathbb {E}} \left[ \underset{(\pmb w_k,\pmb a_k) {\textrm { s.t. (C1)-(C5)}}}{\min}\sum \limits_{k=t+1}^{\infty} \gamma^{k-t}  \left[c_k \left(\pmb a_k,\pmb w_k;\pmb \rho_k,\pmb \lambda_k\right) \Big | {\pmb s}_{t+1} = {\pmb a} \right]  \right] \right \} \\
	\label{value_function}  V\left({\pmb s},{\boldsymbol \theta}\right) :&= \!\!\!\!\!\!\!\!\underset{(\pmb w,\pmb a) {\textrm { s.t. (C1) - (C5)}}}{\mathop{\min}} \left \{ \underset{{\boldsymbol \theta_k}}{\mathbb E} \left[\underset{(\pmb w_k,\pmb a_k) \textrm{ s.t. (C1) - (C5) }}{\min} \left\{\sum \limits_{k=t}^{\infty} \gamma^{k-t} \left[c_k \left(\pmb a_k,\pmb w_k;\pmb \rho_k,\pmb \lambda_k\right) \Big |  {\pmb a}_t = {\pmb a}, {\pmb w}_t = {\pmb w}, {\pmb s}_t = {\pmb s}, {\boldsymbol \theta_t = {\boldsymbol \theta}} \right]\right\}\right] \right\} \\
\label{bellman} 
	{\bar V}(\pmb s) :&= \underset{{\boldsymbol \theta}}{\mathbb E} \left[ V(\pmb s, \boldsymbol \theta) \right] =  \underset{\boldsymbol \theta}{\mathbb E} \; \mathop {\min} \limits_{(\pmb w,\pmb a) {\textrm { s.t. (C1) - (C5)}}} \left\{ c_t(\pmb a, \pmb w;{\boldsymbol \rho},{\boldsymbol \lambda})   + \gamma {\bar V}(\pmb a)  \right\}
	\\ \nonumber	
	\end{align} 
	\hrulefill
\end{figure*}
\begin{align} 
\label{eq.11}
{\cal {\bar C}} := {\mathbb E} \; \left[\sum _{t=0}^{\infty} \sum  _{f=1}^{F} \gamma ^{t} c^f_t \left({\pmb a}^f_t,\pmb{w}^f_t;\pmb{\rho}^f_t,\pmb{\lambda}^f_t\right)\right]
\end{align}
where $\gamma^t$ is an exponentially decaying weight which reduces the impact of distant (and therefore more uncertain) costs, and the expectation is taken with respect to the random variables $\{\pmb r^f_{t},\pmb \rho^f_{t}, \pmb \lambda^f_{t}\}_{f,t}$. This paper assumes these variables are stationary, so that the expectations can practically be estimated. 
In addition the knowledge of the state variables  $\{\pmb r^f_{t},\pmb \rho^f_{t}, \pmb \lambda^f_{t}\}_{f,t}$ is causal, that is, their exact values are revealed at the beginning of every time slot $t$, and all the fetching and caching decisions are made sequentially.   The goal is then to make \textit{sequential} fetching-caching decisions that minimize the expected current plus future costs while satisfying the constraints (C1)-(C5). This gives rise to the following optimization problem 
\begin{align}
{\textrm {(P$1$)} }  \min  \limits_{ \{(\pmb w^f_{k},\pmb a^f_{k}) \}_{f,k\geq t}}    &{\cal {\bar C}}_t:=\sum \limits_{k=t}^{\infty} \sum  \limits_{f=1}^{F} \gamma ^{k-t} {\mathbb E}  \left[c^f_k \left(\pmb a^f_k,\pmb w^f_k;\pmb \rho^f_k,\pmb \lambda^f_k\right)\right]  \nonumber\\
\mathrm{s.t.}\;\;\;\; & \textrm{(C1) -- (C5)}, \nonumber
\end{align}
Due to~(C1), (P$1$) is indeed a DP, because current decisions will influence not only current, but also future costs. As a result, one needs to~\cite[p.~10]{RL}: (i) identify the current and the expected future aggregated cost (the latter gives rise to the so-called value function), (ii) write the Bellman optimality conditions; and (iii) suggest a method to  estimate the value function. In addition, note that (P$1$) is indeed decomposable across files, since the fetching and caching prices are different for $f$, $t$, and $m$. In most practical setups, those prices are in fact (stochastic) Lagrange multipliers associated with constraints that have been handled using dual decomposition techniques (see, e.g.,~\cite{Dual}). Definition of those prices for the investigated caching setup is certainly of interest, but out of the scope of this paper. For this reason, in the sequel the focus is on solving (P$1$) for a single file $f$. Hence, the superscript $f$ will be dropped. In addition, due to stationarity the subscript $t$ will be dropped occasionally as well. 
\vspace{+.1 in}
\section{DP-based optimal fetching and caching}
To solve (P$1$), collect all the storage state variables in vector ${\pmb s}_t$, define $\boldsymbol \theta_{t} =[{\pmb r}_{t},{\pmb \rho}_{t},{\pmb \lambda_{t}}]$, and assume that their initial values ${\pmb s}_{0}$ and ${\pmb \theta}_0$ are given. Then the optimal fetch-cache decisions  $({\pmb w}^{\ast}_{t},{\pmb a}^{\ast}_{t})$ are expressible  as the  solution to \eqref{long1}. The objective in \eqref{long1} is then rewritten in \eqref{long2} as the summation of current and discounted average future costs. The form of \eqref{long2} is testament to the fact that problem (P$1$) is a DP and the caching decisions $\pmb a$ influence not only the current cost $c_{t}(\cdot)$ but also future costs through the second term as well. Bellman equations can be  leveraged  for tackling such a DP. Under the stationarity assumption for variables $\{\pmb \theta_{t}\}_{t}$, the term accounting for the future cost can be rewritten in terms of the \textit{ stationary value function} $V\left(\pmb s, \boldsymbol \theta \right)$ \cite[p.~68]{RL}. This function, formally defined in \eqref{value_function}, captures the minimum aggregated cost starting from ``state''~$(\pmb s , \boldsymbol \theta)$. Assuming that prices and requests are i.i.d. across time slots and nodes, it can be shown that the solution for (P$1$) can also be found in terms of the \textit{reduced value function},  ${\bar V}(\pmb s) := \underset{{\boldsymbol \theta}}{\mathbb E} \left[ V(\pmb s, \boldsymbol \theta) \right]$, which depends only on~$\pmb s$~\cite{Antonio}. The reduced value function is important since not only enables solving (P$1$), but also alleviates the computationally burden of estimating $V\left(\pmb s, \boldsymbol \theta \right)$, which has a much larger dimensionality. 

By rewriting the reduced value function 
recursively as the summation of the instantaneous cost and the aggregated future values, one readily arrives at the Bellman equations
provided in (11). Thus, the problem at time $t$ reduces to 
\begin{align}
{\textrm {(P$2$)} }\quad  \min \limits_{(\pmb w, \pmb a)} \;\;\;\;&  c_t(\pmb a, \pmb w;\pmb \rho_t,\pmb \lambda_t)   + \gamma {\bar V} \left(\pmb a\right)  \nonumber\\
\mathrm{s.t.}\;\;\;\; &\textrm{(C1)- (C5)}, \nonumber
\end{align}
which requires the estimation of the reduced value function $\bar V (\cdot)$. This can be done using the value iteration algorithm as tabulated in Alg. 1; for a detailed
discussion cf.~\cite[p.~100]{RL}. Alternatively, if the distributions of the random parameters are \textit{unknown}, stochastic online solvers based on $Q$-learning are well motivated~\cite[p. 148]{RL}. This is left as future work. 
\begin{figure*}[t]
	\centering
	\vspace{-.025cm}
	\begin{subfigure}{.32\textwidth}
		\centering
		\includegraphics[width=\textwidth]{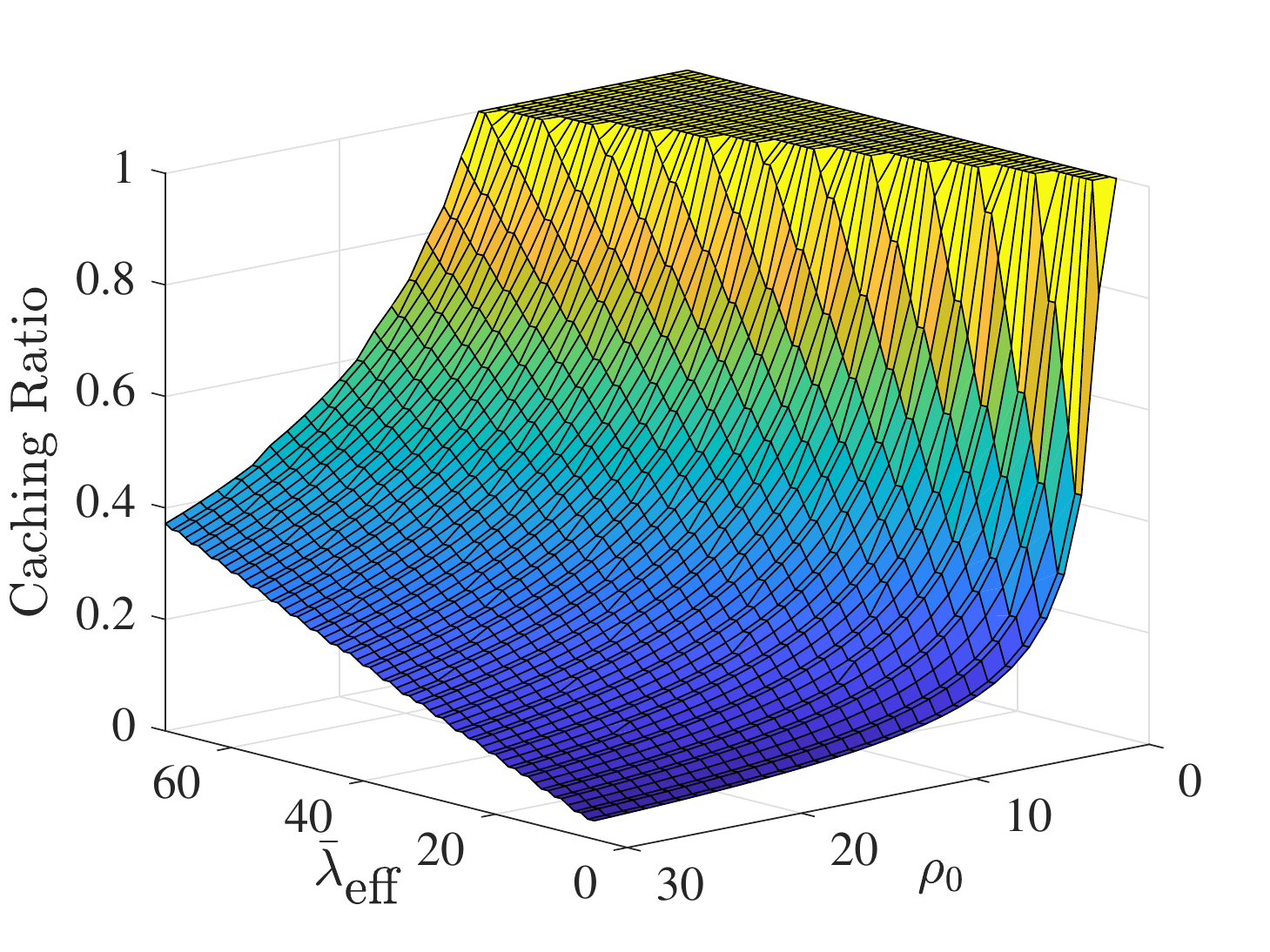}
		\caption{}
		\label{result1}
	\end{subfigure}%
	\hspace{0.1cm}
	\begin{subfigure}{.32\textwidth}
		\centering
		\includegraphics[width=\textwidth]{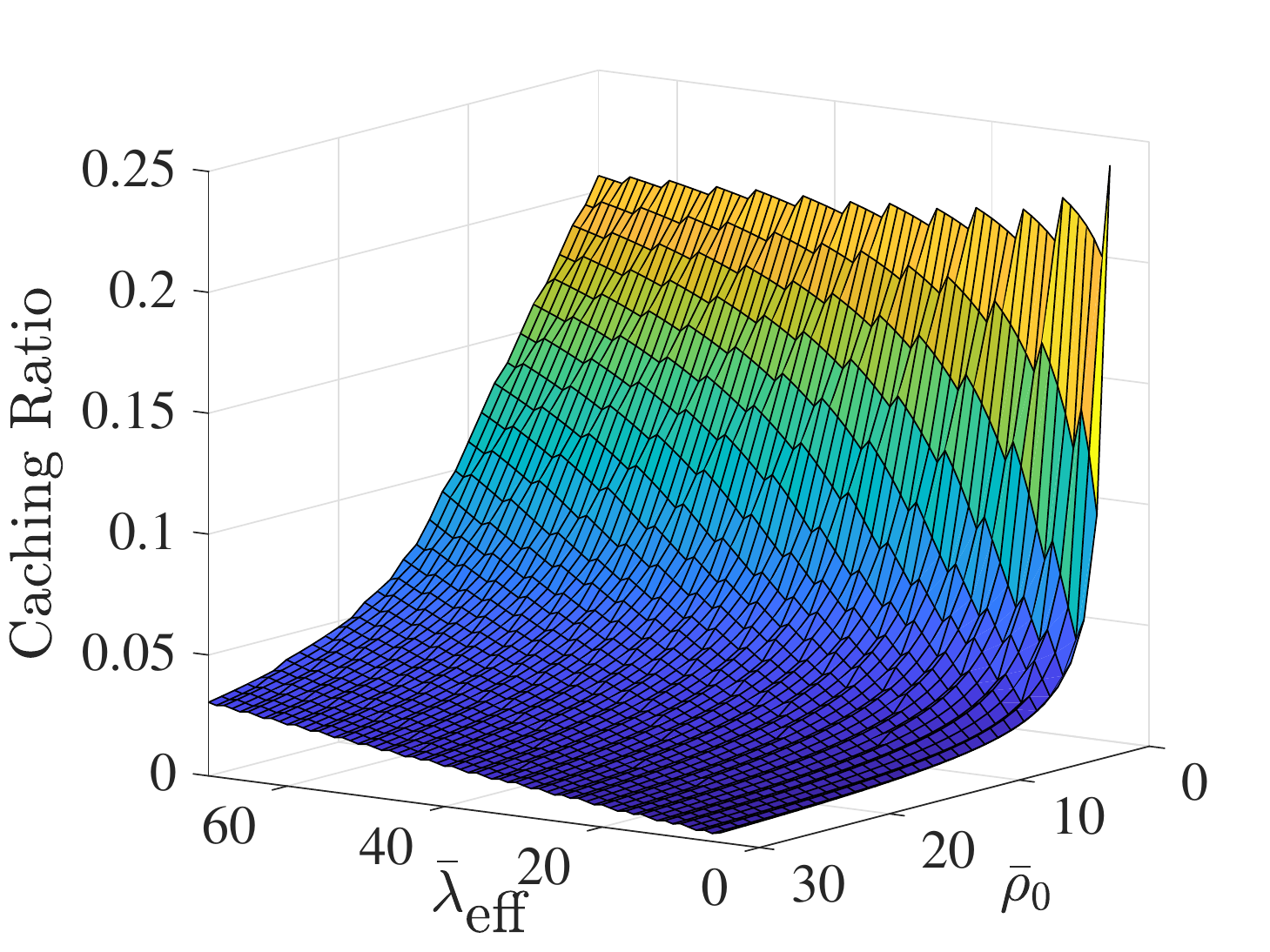}
		\caption{}
		\label{result2}
	\end{subfigure}%
	\hspace{0.1cm}
	\begin{subfigure}{.32\textwidth}
		\centering
		\includegraphics[width=\textwidth]{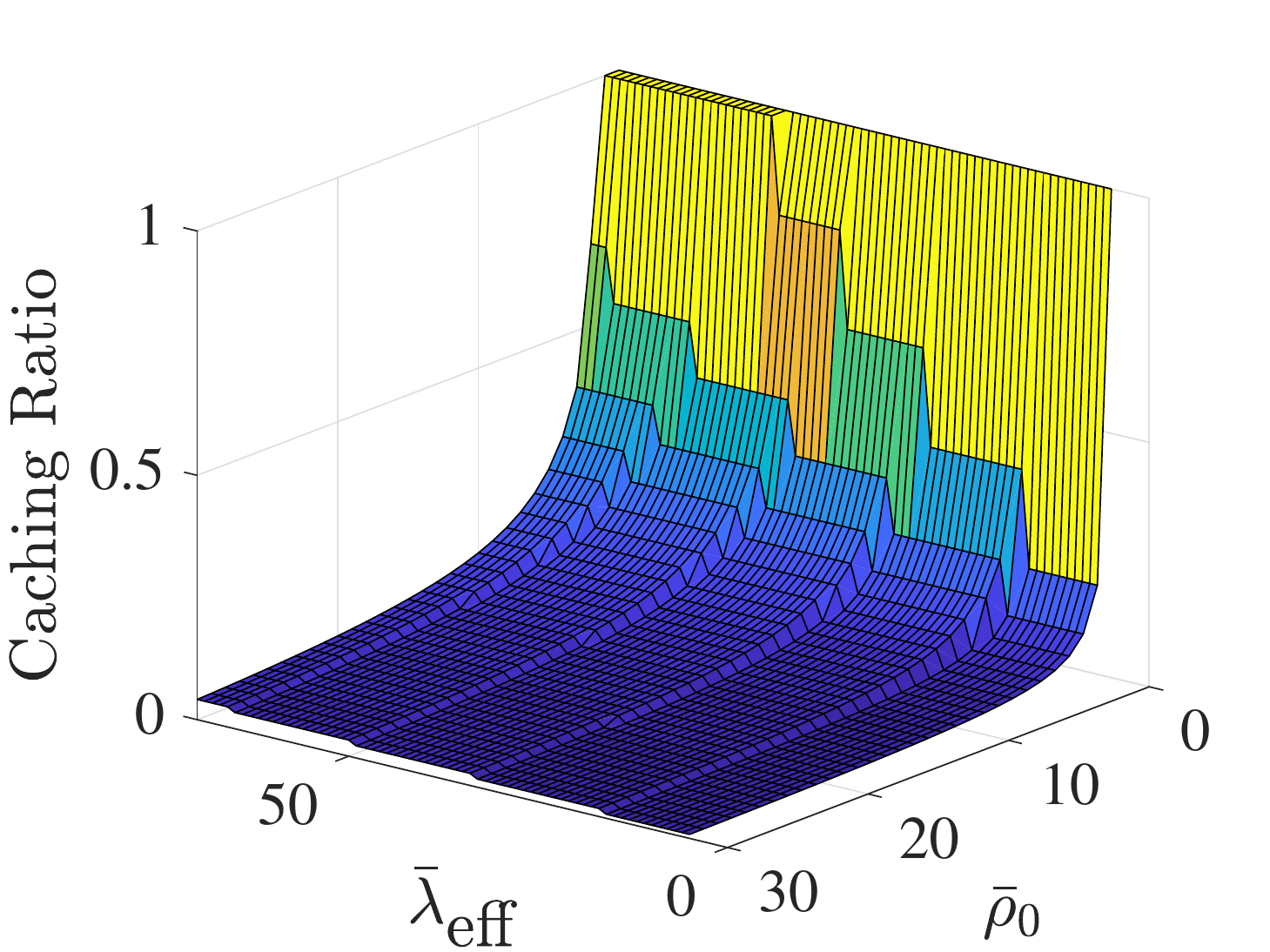}
		\caption{}
		\label{result4}
	\end{subfigure}
	\begin{subfigure}{.32\textwidth}
		\centering
		\includegraphics[width=\textwidth]{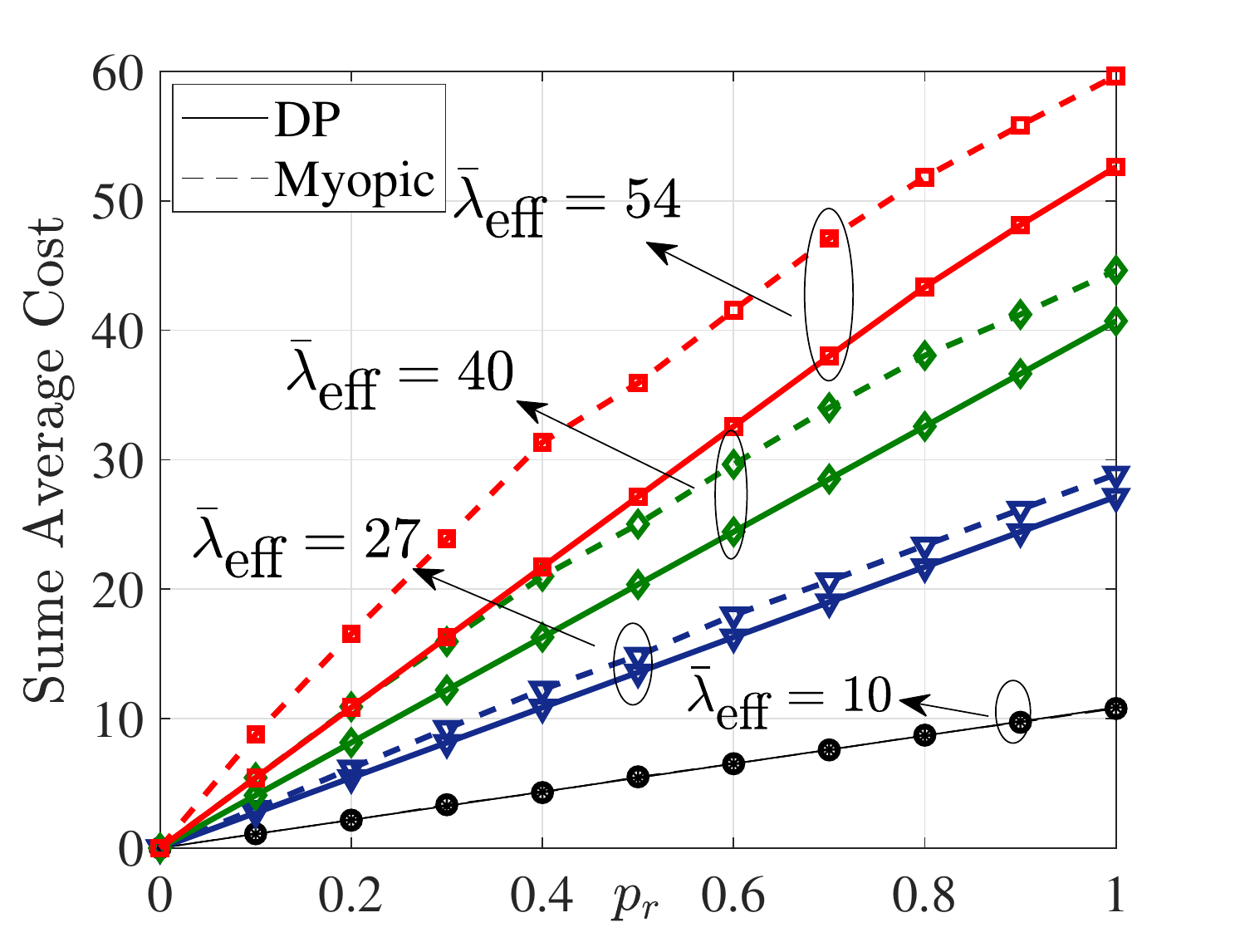}
		\caption{}
		\label{result5}
	\end{subfigure}
	\begin{subfigure}{.32\textwidth}
		\centering
		\includegraphics[width=\textwidth]{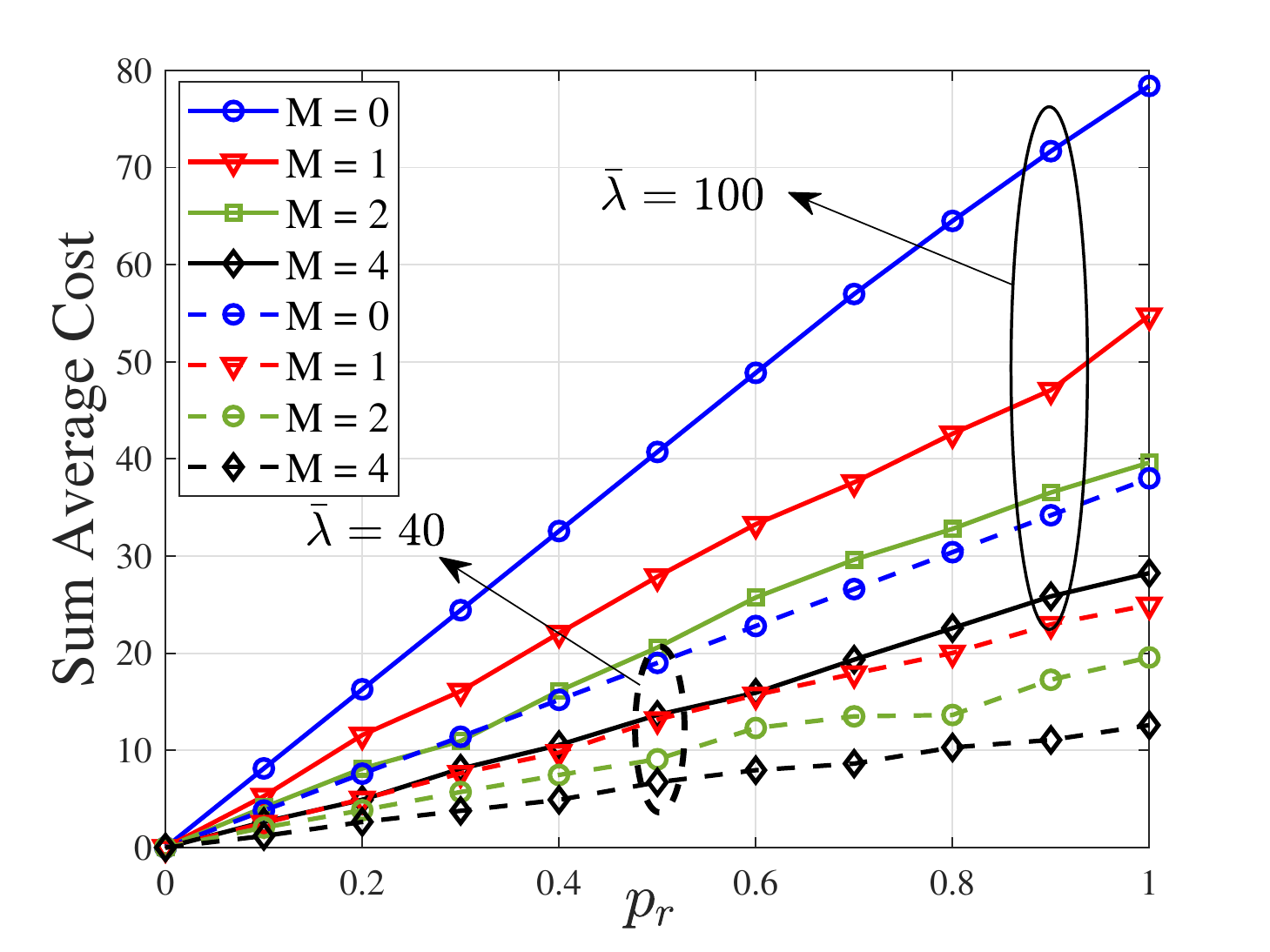}
		\caption{}
		\label{result6}
	\end{subfigure}
	\vspace{-0.1in}
	\caption{\small Caching ratio versus (${\bar \lambda_{\textrm{eff}}},{\bar \rho_0}$) for different initial states $s_0$,$r_0$, and request probability $p_r$;  (a) $s_0 = 1$, $r_0 = 1$, and $p_r = 0.5$; (b) $s_0 = 0$, $r_0 = 0$, and $p_r = 0.5$; and (c) $s_0 = 0$, $r_0 = 0$, and $p_r = 0.05$. (d) Performance of DP   versus myopic approach for different $\bar \lambda_\textrm {eff}$. Performance of of proposed algorithm for different number of regular nodes.}
	\vspace{-0.0 in}
	\label{F:num_exp}
\end{figure*}
\vspace{.1cm}
\subsection{Discussing the solution to (P2)} To gain insights on the optimal fetching-caching solution, suppose that at time $t$, we have $s_{t,0}=0$ and the CN receives file request from node $m'$. In such a case, $r_{t,0}=1$ and thus (C$2$) requires $w_{t,0} + \sum_{m \in {\cal M}_t}  {\underline w}_{t,m}\geq 1$, where ${\cal M}_t$ denotes the set of nodes for which $s_{t,m}=1$. Since the reduced value function does not depend on the fetching decision ${\pmb w}_t$ and the cost $c_t$ is linear in ${\underline w}_{t,m}$ [cf. \eqref{cost} and \eqref{c_0}], we have that the optimal fetching is
$${\underline w}_{t,m}^* = \left\{
\begin{array}{ll}
	1 & \quad \text{if~} m=\arg\min_{m''\in {{\cal M}_t \cup \left\{{\textrm {Cloud}}\right\}}} {\underline  \lambda}_{t,m''} \\
	0 & \quad \text{for~ all~ other} ~m
\end{array}
\right.$$
Deciding on the caching variable ${\pmb a}_t$ is more complicated. Suppose first that the value function can be written as $\bar V (\pmb s) = \sum_{m=0}^{M} \bar V_{m} (s_m)$ and that node $m'$ was the one requesting the content. Then, we have that, the cost of setting $a_{t,m}=1$ is
$$c_{\{a_{t,m}=1\}} = \left\{
\begin{array}{ll}
\rho_{t,m} + \gamma {\bar V}_m(1) & \quad \text{for~} m\in {\cal M}_t^+ \\
\rho_{t,m} + {\overline \lambda}_{t,m} + \gamma {\bar V}_m(1) & \quad \text{for~ all~ other} ~m
\end{array}
\right.$$
where ${\cal M}_t^+={\cal M}_t \cup \{0,m'\}$. 
On the other hand, the cost of setting $a_{t,m}=0$ is just $c_{\{a_{t,m}=0\}}=\gamma {\bar V}_m(0)$ for all~$m$. Hence, to obtain the optimal ${\pmb a}_t$ one only needs to solve \linebreak $(M+1)$ separate tests of the form $c_{\{a_{t,m}=1\}} \gtrless c_{\{a_{t,m}=0\}}$, one per node. However, in most cases the assumption of $\bar V (\pmb s) = \sum_{m=0}^{M} \bar V_{m} (s_m)$ does not hold and, as a result, caching decisions are coupled across nodes. In addition, while for a small number of users (say 10 or less) both the estimation of ${\bar V}(\pmb s)$ and the computation of the multiple decision rule is affordable, the computational burden grows exponentially as $M$ increases. As a result, in networks with a large number of users, alternative schemes that try to impose additional structure on ${\bar V}(\pmb s)$ are well motivated.

\begin{algorithm}[t]
	\SetKwInOut{Input}{Input}
	\SetKwInOut{Output}{Output}
	\underline{Set} $\bar V(\pmb s) = 0, {\textrm {for }} \forall \pmb s \in {\mathcal S}$; $\mathcal S := \left\{ {\pmb v} \big| \pmb v \in \left\{0,1\right\}^{M+1} \right\} $
	
	\Input{$\gamma<1$, pdf for $\pmb \rho,{\pmb \lambda}$ and $\pmb r$}
	\Output{$\bar V(\cdot)$}
	\While {$|\bar V^i (\pmb s)-\bar V^{i+1} (\pmb s)| < \epsilon; \forall \pmb s \in {\mathcal S}$} { \For{  $\forall \pmb s \in {\mathcal S}$} {$\bar{V}^{i+1}(\pmb s) = \mathbb{E} \; \mathop {\min} \limits_{( {\pmb \omega},\pmb \alpha) {\textrm { s.t. (C1)-(C5) }}} \left\{  c(\pmb \alpha ,{\pmb \omega} ; {\pmb \rho}, {\pmb \lambda}) + \gamma {\bar V^i(\pmb \alpha)}\right\}$}
		$i = i+1$}    
	\caption{Value iteration  for finding $\bar V \left(\cdot\right)$}
\end{algorithm}
\vspace{+.15 in}
\section{Numerical tests}
Here we run numerical simulations to assess the behavior of the developed caching policy. Due to space limitations, the focus will be on the behavior of the CN, but the analysis can be generalized to nodes with $m\geq 1$ as well. The caching cost $\rho_0$ is considered to be uniformly drawn from $[0, 2 \bar \rho_0]$. In addition, to model the fetching costs over the entire network, we define the ``effective'' fetching cost $\lambda_{\textrm{eff}}$ as the cost of fetching from any node or cloud to the CN. $\lambda_{\textrm{eff}}$ is also considered to be uniformly drawn from $[0, 2 \bar \lambda_{\textrm{eff}}]$. To assess the CN's policy, let us define the {\it caching ratio} as the ratio of the number of caching decisions divided by the total number of decisions. Fig~\ref{result1}. depicts this ratio for different values of $\bar \lambda_{\textrm{eff}}$ and $\bar \rho_0$, for $s_0 = 1$, and $r_0 = 1$. The requests from the CN are assumed to follow a Bernoulli distribution with $p_r = 0.5$. The plot reveals a non-symmetric cache-versus-fetch trade off curve for the CN's policy, showing clearly how caching is preferred over fetching as the fetching cost increases.  For small enough caching costs, the ratio goes to one (i.e., caching happens with probability $1$). Changing the initial state to $s_0 = 0$, and $r_0 = 0$ clearly affects the CN's policy as depicted in Fig.~\ref{result2}. Now, even when caching is cheap, caching decisions are rare (always below $20$ \%). Intuitively, this is due to the fact that the file has not already been requested ($r_0 = 0$) and, thus, caching is only made to take advantage of low fetching costs. Fig.~\ref{result4} shows the caching ratio for $s_0 = 1$, $r_0 = 1$, and $p_r = 0.05$. The results confirm the relevance of the popularity of a content, since for a wide range of caching costs fetching is always decided. Interestingly, even when fetching is costly, the algorithm tends to fetch. The reason is that, due to its low popularity, the file will have to remain for a long time in the memory before being used, entailing a very high \textit{aggregated} caching cost. 

{ We now compare the performance of the proposed method with that of simpler schemes. Specifically,} Fig.~\ref{result5} reports the sum average cost of the proposed DP approach compared with that of a myopic one. The myopic policy stores a file only if the file is requested or is locally stored and the current caching cost is less than the fetching one, i.e., $\rho_{0,t} < \lambda_{\textrm{eff},t}$. Here, $\bar \rho_{0} = 10$, and as this result demonstrates the DP approach outperforms the myopic one as the difference between average costs increases. This is expected because in the myopic approach the file will be stored whenever the fetching is costly; however, the DP-based approach will also consider the probability of the file being requested in future instants.  {The next goal is to assess  the benefits associated with distributed caching by comparing the obtained cost with the cost incurred when only the CN is equipped with a cache. To that end, Fig.~\ref{result6} reports the results for different number of caching nodes  $M \in \left\{0, 1, 2, 4\right\}$. The fetching costs over any link of the network are assumed to be i.i.d, and uniformly drawn from $[0-2 \bar \lambda]$. The caching costs are also considered i.i.d and drawn from $[0-2 \bar \rho]$. This plot depicts the sum average cost incurred by the CN for $\bar \rho = 62$, two values of $\bar \lambda$ ($40$ and $100$),  and different probabilities of requesting the files. Clearly, increasing $M$ allows the CN to meet the demands and, hence, reduces the aggregated cost.  As expected, this reduction saturates as $M$ increases. All in all, the results corroborate the benefits of having a system equipped with distributed caching infrastructure. } 
\vspace{-.1 in}
\section{Conclusion}
This paper illustrated how decomposition and dynamic programming tools can be used to design optimal fetching-caching schemes for a network of caches. In particular, we focused on an architecture  formed by a central controller connected to the cloud and to a set of local nodes. At each time instant the controller received different content requests and had to decide: i) if the content was going to be downloaded from the cloud or from the local nodes (provided that it was available locally), and ii) if some of the local nodes should cache the content in their local memory for future use. Upon defining suitable fetching and caching costs (that varied across time, nodes and contents), the problem was formulated as a dynamic program and its optimal solution was obtained using different strategies to reduce the associated computational burden.

\newpage
\bibliographystyle{IEEEbib}
\bibliography{biblio}
\end{document}